\documentclass[
  aps,
  prb,
  twoside,
  twocolumn,
  showpacs,
  floatfix,
  superscriptaddress,
  10pt,
  preprintnumbers,
  citeautoscript,
]{revtex4}

\usepackage{amsmath}
\usepackage{comment}
\usepackage{graphicx}
\usepackage[utf8]{inputenc}
\usepackage{times}
\usepackage{physics}
\usepackage{amssymb}

\begin{document}

\title{Investigation of oxygen-vacancy complexes in diamond by means of \textit{ab initio} calculations}

\newcommand{\Wigner}{Wigner Research Centre for Physics, P.O. Box 49, H-1525 Budapest, Hungary}
\newcommand{\BME}{Department of Atomic Physics, Institute of Physics, Budapest University of Technology and Economics, M\H{u}egyetem rakpart 3., H-1111 Budapest, Hungary}

\author{Nima Ghafari Cherati}
\affiliation{\Wigner}
\affiliation{\BME}
\author{Gerg\H{o}  Thiering}
\affiliation{\Wigner}

\author{\'Ad\'am Gali}
\affiliation{\Wigner}
\affiliation{\BME}



%
\begin{abstract}
Point defects in diamond may act as quantum bits. Recently, oxygen-vacancy related defects have been proposed to the origin of the so-called ST1 color center in diamond that can realize a long-living solid-state quantum memory. Motivated by this proposal we systematically investigate oxygen-vacancy complexes in diamond by means of first principles density functional theory calculations. We find that all the considered oxygen-vacancy defects have a high-spin ground state in their neutral charge state, which disregards them as an origin for the ST1 color center. We identify a high-spin metastable oxygen-vacancy complex and characterize their magneto-optical properties for identification in future experiments.
\end{abstract}

\maketitle
\section{Introduction}
\label{sec:introduction}
In quantum technology, quantum information processing by diamond color centers operating at room temperature has attracted a great attention~\cite{Wrachtrup2001, Thiering2020, Zhang2020}. More than 500 color centers in diamond have been reported to date~\cite{Zaitsev2000} but only few of them have been unambiguously identified as single-photon emitters or quantum bits. The most studied color center acting as quantum bit is the negatively charged nitrogen-vacancy (NV$^-$) center in diamond, which has been successfully employed in quantum technologies~\cite{Gruber1997, Kurtsiefer2000NV, Maze-gali2011NV, DOHERTY2013, Gali2019}. Some properties of NV$^-$ are not favorable, which has generated a large interest in searching for alternative candidates with improved parameters for a given application~\cite{Weber2010, Zhang2020, Gali2023}.

The so-called ST1 color center as a single photon emitter was discovered in diamond nanowire samples~\cite{Lee2013} that were later accidentally found in chemical vapor deposited (CVD) diamond too~\cite{John2017, Balasubraman2019}. 
The center exhibits a relatively sharp zero-phonon-line (ZPL) with coherent emission at around $550$~nm ($2.25$~eV) and a wide phonon sideband stretching up to $750$~nm. They found an optically detected magnetic resonance signal (ODMR) associated with this color center, which was assigned to a metastable triplet state of the defect. The fluorescence spectrum and the zero-field-splitting parameters of the triplet state ($D=1135$~MHz, $E=139$~MHz) provide no direct information about the chemical composition of the defect. The spectroscopy data implies that the defect has relatively low but not trivial (C$_1$) symmetry because of the selection rules manifested in the spin-dependent intersystem crossing, and the symmetry axis should be along [110] plane.

In a recent report, a systematic study has been carried out by implanting different ions into chemical vapor deposited (CVD) diamond, and then the sample was imaged by scanning confocal microscope to monitor the presence of ST1 color center~\cite{Luhmann2022}. They found that the occurrence of ST1 color center is associated with the implanted oxygen ions, thus the defect should contain oxygen~\cite{Luhmann2022}. Because of the low yield of the ST1 center, it was speculated that the ST1 center should be a complex of oxygen and vacancy, and it might also contain another impurity such as hydrogen~\cite{Luhmann2022}. We note that $^1$H isotope has almost 100\% abundance, thus it should be manifested in the ODMR signal due to the hyperfine interaction between the $^1$H nuclear spin and the electron spin. It was speculated that some complex of vacancy and oxygen separated at certain distances for providing relatively small $D=1135$~MHz spin-spin interaction could be also the origin of the ST1 center~\cite{Luhmann2022}.

It is little known about oxygen in diamond. Two electron spin resonance (ESR) centers were associated with vacancy-oxygen complexes in diamond: WAR5 center~\cite{Hartland2014} and "OVH" center~\cite{Breeze2016} were assigned to oxygen-vacancy and oxygen-vacancy-hydrogen complexes, respectively. We note that the neutral oxygen-vacancy center (OV$^0$) is isovalent to the NV$^-$ and it was speculated that the OV$^0$ center might have the same optical activity and properties like the NV$^-$ center in diamond. However, previous first principles calculations showed~\cite{Thiering2016} that the ground state of the OV$^0$ center has indeed similar ground state and spin properties to those of the NV$^-$ center in diamond but the excited states differ in the two systems. A fast nonradiative decay was found from the optical excited state to the ground state of OV$^0$ defect~\cite{Thiering2016}. The high-spin $S=1$ ground state and the lack of fluorescence disregard OV$^0$ as the origin of the ST1 defect. Nevertheless, it cannot be excluded that other complexes than OV defect might be developed between oxygen and vacancy, which should be explored.

Here we carry out a systematic study of complexes of oxygen and vacancies in diamond by means of density functional theory calculations. The paper is organized as follows. In Sec.~\ref{sec:methodology} we briefly describe
the applied methodology. In Sec.~\ref{sec:results} we discuss the results of atomic
simulations on the complexes of oxygen and vacancy with considering an increasing separation distance between the oxygen atom and the vacancy in the diamond lattice. We determined the relative stability, electronic structure, and charge transition levels of these complexes. For the relevant complex we carry out a detailed study by computing its magneto-optical parameters. Finally, we conclude our paper in Sec.~\ref{sec:conclusion}.

\section{Methodology}
\label{sec:methodology}

We applied Kohn–Sham density functional theory (DFT) approach in our calculation, which is implemented in Vienna \textit{ab initio} simulation package (VASP)~\cite{kres1} within projector-augmented-wave (PAW)~\cite{PAW1994} method. Plane wave basis with a cutoff energy of 370~eV was employed. 

We used a screened hybrid density functional, the Heyd, Scuseria, and Ernzerhof (HSE06)~\cite{Paier2006HSE}, which can reproduce the experimental band gap and defect levels in the gap with 0.1~eV accuracy~\cite{Deak2010}.

In order to model defects, we built a $4\times4\times4$ simple cubic 512-atom supercell from the conventional unit cell. With this size of supercell, accurate results can be achieved with sampling the Brillouin zone only at the $\Gamma$ point. We applied 10$^{-3}$~eV/\AA\ threshold for ionic forces in the geometry optimization procedure in the HSE06 calculations. We also applied a computationally less demanding Perdew-Burke-Ernzerhof (PBE) functional~\cite{PerdewPBE1996} for calculating the vibrational modes. In this case, the ionic force threshold was set to a stringent value at 10$^{-4}$~eV/\AA.
The optimized lattice constants are $a=3.565$~{\AA}
and $a=3.544$~{\AA} for PBE and HSE06 functionals, respectively. In the geometry optimization procedure the global energy minimum of the adiabatic potential energy surface (APES) is obtained. 

In order to verify the stability of defects at zero kelvin temperature, we calculate the formation energy $E^q_\text{f}$ of defects in charge state $q$, which is defined~\cite{Northrup1996} as 
\begin{equation}
\label{eq:Eform}
    E^q_\text{f} = E^q_\text{tot} - \sum_{i} n_i \mu_i + qE_\text{Fermi} + E^q_\text{corr},
\end{equation}
where $ E^q_\text{tot}$ is the total energy of defective supercell, and $ \mu_i$ is the corresponding chemical potential for atom type type $i$.
$E_\text{Fermi}$ is the Fermi-level referenced to the valence band maximum (VBM) as calculated in the perfect supercell. $ E^q_\text{corr}$ is the correction of the total energy for the charged supercells.
For calculating the charge correction term we used Lany and Zunger correction~\cite{Makov1995, luny2008}.

In our study, defective supercells contain carbon and oxygen atoms. The chemical potential of carbon can be calculated from diamond as a reference. The challenging part is to define oxygen chemical potential since oxygen and carbon create a gas phase at ambient conditions (e.g., see the discussion in Ref.~\cite{gali2007}). Furthermore, we assume that oxygen is introduced by implantation to diamond which is not an equilibrium process. We decided to choose CO$_2$ molecule at zero kelvin as a reference for the chemical potential of oxygen. In this small molecule, the zero-point-energy correction coming from the vibrations cannot be neglected, so we added this correction (half of sum of individual oscillator energies) to the electronic total energy of CO$_2$ that was modeled in a large simulation box with the edge of 15~\AA .

An important characteristic magneto-optical parameter is the so-called zero-field-splitting (ZFS) for high-spin ($S>1/2$) defects as we introduce below. In the triplet state, the three spin levels ($m_S=\{0;+1;-1\}$) in the absence of external magnetic fields are degenerate in an isotropic environment. However, the crystal field of the defective crystal introduce spatial anisotropy which may split these degenerate levels. By preserving axial symmetry in the crystal field, two levels, $m_S=0$ and $m_S=\{+1;-1\}$, are split by the zero-field-splitting parameter $D$. If the axial symmetry was also broken then three levels appear separated by $D-E$ and $D+E$ energies, where $E$ is the orthorhombic ZFS parameter. The corresponding spin-Hamiltonian is written as
\begin{equation}
   \hat{H}= D \bigg(S_z^2 - \frac{1}{3} S(S+1)\bigg) + E ( S_x^2 - S_y^2 )\text{,}
\end{equation} 
where  $S=1$ is the total spin, and $S_{x,y,z}$ are the spin components. The origin of the ZFS could be spin-orbit interaction or electron spin dipole-dipole interaction. Since the symmetry of the considered defects is low and no heavy elements are involved, we only consider the electron spin dipole-dipole interaction as a source of ZFS as implemented by Martijn Marsman in VASP within the PAW formalism~\cite{Bodrog2013, Ivady2018, Gali2019}.

\begin{figure*}[ht!]
    \centering
    \includegraphics[scale=0.1]{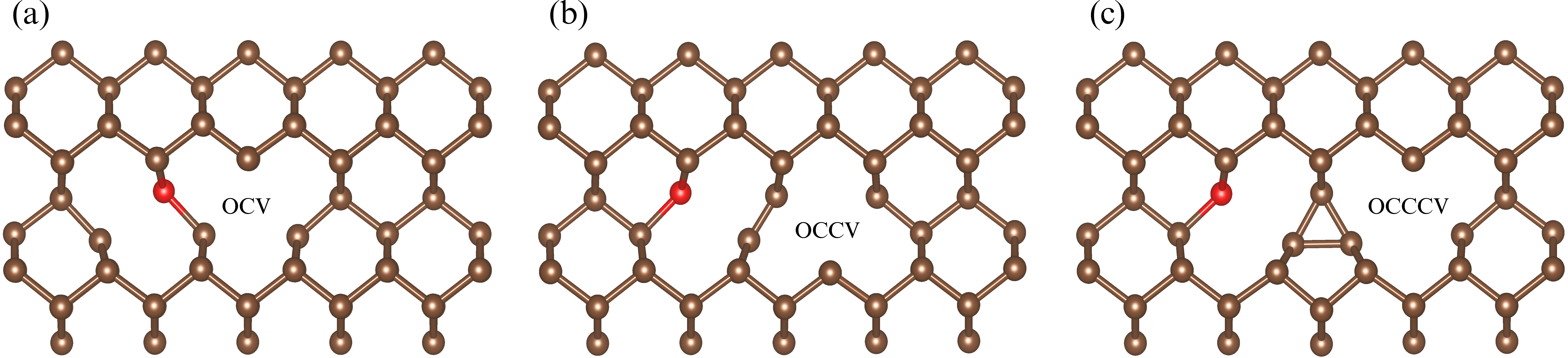}
    \caption{\label{fig:structures} Geometry of the defects after optimization along [110] mirror plane for (a) OCV, (b) OCCV, and (c) OCCCV. O and C atoms are depicted as red and brown spheres, respectively.}
\end{figure*}

The hyperfine tensors of the paramagnetic defects were also calculated. With considering the relativistic effects, hyperfine tensor of the given nucleus consists of the Fermi-contact term and the dipole-dipole term, which are given as
\begin{equation}
\label{eq:hyper}
   A^I_{ij} = \frac{1}{2S}\int n_s(r) \gamma_I \gamma_e \hbar^2 \biggr[ \Bigl( \frac{8 \pi}{3} \delta(r) \Bigl) + \Bigl( \frac{3x_ix_j}{r^5} - \frac{\delta_{ij}}{r^3} \Bigl) \biggr] d^3r \text{,}
\end{equation}
where $n_s(r)$ is the spin density of the spin state $S$, $\gamma_I$ is the nuclear Bohr magneton of nucleus $I$, and $\gamma_e$ is the electron
Bohr magneton. The Fermi-contact term is proportional to the spin density localized at the site of the nucleus, which
is dominant compared to the dipole-dipole term. We calculate the hyperfine tensor and diagonalize it in order to obtain hyperfine constants.
The hyperfine constants can be directly compared to experimental data. The ratio of the Fermi-contact and dipole-dipole terms characterizes the shape of the spin density.

The fluorescence spectrum of the defects was computed within the Huang-Rhys (HR) theory~\cite{Huang1950}. This requires to determine the APES of the electronic excited state as well as the phonons that may participate in the optical transitions. The excited state of the defect was calculated by $\Delta$SCF method~\cite{Gali2009}. The forces in the excited state can straightforwardly be calculated within $\Delta$SCF method where the global energy minimum of the APES in the electronic excited state can be determined. With obtaining the global energy minimum in the ground state (GS) energy $E_\text{g}$ and excited state (ES) energy $E_\text{e}$ the ZPL energy was calculated as $E_\text{ZPL}=E_\text{e}-E_\text{g}$.
The phonon sideband of the fluorescence spectrum as a function of the photon energy $\hbar \omega$ and temperature is expressed~\cite{Alkauskas2014,Jin2021,Shang2020} as
\begin{align}
\label{eq:pl}
L(\hbar \omega, T) &=  \frac{n \omega^3}{3\pi \epsilon_0 c^3 \hbar} \big|\mu_\text{eg}\big| \sum_i\sum_j F_{\text{e}:j}(T) \;   \abs{
\braket{\psi_{\text{e}:j}}{\psi_{\text{g}:i}}}^2 \\ \nonumber
  & \cross \delta (E_\text{ZPL} +E_{\text{e}:j} - E_{\text{g}:i} - \hbar \omega ) \text{,} 
\end{align}
where  $n$ is the refractive index of the material, $\mu_\text{eg}$ is the electronic transition dipole moment, and $\ket{\psi_{\text{g}:i}}$ ( $\ket{\psi_{\text{e}:j}}$) is the $i$th
($j$th) wavefunction of the system in the ground state (excited state) with vibrational energy  $E_{\text{g}:i}$ ($E_{\text{e}:j}$). The thermal distribution function of the vibrational energy in the excited state is
\begin{equation}
    F_{\text{e}:j}(T) = \frac{e^{-\frac{E_{\text{e}:j}}{k_\text{B}T}}}{\sum_j e^{-\frac{E_{\text{e}:j}}{k_\text{B}T}}} \text{,}
\end{equation}
where $k_\text{B}$ is the Boltzmann constant. In solids within a harmonic approximation, we express the nuclear wave function as a product of a vibrational wave functions,
\begin{equation}
\ket{\psi_{\text{g}:i}} = \Pi_k  \ket{\phi_{kn_k^{\text{g}:i}}} , \, \,  \ket{\psi_{\text{e}:j}} = \Pi_k  \ket{\phi_{kn_k^{\text{e}:j}}} \text{,}
 \end{equation}
 where $n_k^{\text{g}:i}$ ($n_k^{\text{e}:j}$) is the number of $k$th phonon in the $i$th ($j$th) vibrational state of the GS (ES), and $\ket{\phi_{kn}}$ is the $n$th vibrational state of the GS (ES).
The vibrational energies for the GS and ES are
\begin{equation}
E_{\text{g}:i} = \sum_k n_k^{\text{g}:i} \hbar \omega_k^\text{g} \,\, \text{and} \,\, E_{\text{e}:j} = \sum_k n_k^{\text{e}:j} \hbar \omega_k^\text{e} \text{,}
 \end{equation}
 respectively.
Evaluating the multidimensional overlap integrals (${\braket{\psi_{\text{e}:j}}{\psi_{\text{g}:i}}}$) in Eq.~\ref{eq:pl} is very challenging. Considering the HR theory~\cite{Huang1950}, assuming that the potential energy surfaces of the ES and the GS are identical except for fixed displacement due to the difference in their equilibrium structures, i.e., $ \omega_k^\text{g} =  \omega_k^\text{e}$, it simplifies the equation.
 
With the use of the HR theory and generating function approach~\cite{kubo1995}, Eq.~\ref{eq:pl} can be expressed as the Fourier transform of the generating function $G (t,T)$,
\begin{equation}
\label{eq:pl_T}
L(\hbar \omega, T) \propto  \omega^3 \int_{-\infty}^{{+\infty}} 
G(t,T) e^{i \omega t- \frac{\gamma}{\hbar} \abs{t}-i\frac{E_\text{ZPL}}{\hbar}t} \text{,}
\end{equation}
where $\gamma$ accounts for the broadening of the lineshape and $G (t,T)$ is expressed as
\begin{equation}
\label{eq:generationF}
G(t,T) = e^{S(t)-S(0)+C(t,T)+C(-t,T)-2C(0,T)} \text{,} 
\end{equation}
where $S(t) = \sum_s S_k e^{i \omega t} $ that $(S_k)$ is the partial HR factor and accounts for the average number of $k$th phonons that participate in the transition. The $C(t,T)= \sum_s \Bar{n}_k(T) S_k e^{i \omega t} $, where $\Bar{n}_k(T)$ is
the average occupation number of the $k$th phonon mode and was given as
\begin{equation}
    \Bar{n}_k(T) = \frac{1}{e^{\frac{\hbar \omega}{k_B T}-1}} \text{.}
\end{equation}

For evaluation of Eq.~\ref{eq:pl_T}, with having the ZPL energy and phonon frequencies in hand,  the partial HR factor $(S_k)$, can be written as
\begin{equation}
    S_k = \frac{1}{2\hbar} \omega_k \Delta Q^2 \text{,}
\end{equation}
where $\Delta Q$ is the mass-weighted difference of the ground state and the excited state geometries, evaluated as 
\begin{equation}
\label{eq:deltaq}
\Delta Q = \sum_{i,j} m_i\; \Delta R^2_{ij} \text{,}
\end{equation}
where $i$ is the index of the atom, $j= \{x,y,z \}$, $m_i$ is atomic mass of atom $i$, and $\Delta R = R_{\text{e}:ij} - R_{\text{g}:ij}$ is the distortion vector while $R_{\text{e}:ij}$ and $R_{\text{g}:ij}$ is the atomic coordinate in the electronic excited state and ground state, respectively. $\Delta Q$ determines
the phonon sideband in the photoluminescence spectrum for the optical transition.
The HR factor is related to Debye-Waller factor, which quantifies the ratio of zero-phonon or no-phonon emission and the total emission involving phonons that can be directly observed in the fluorescence spectrum. The DW factor ($W$) can be calculated as $W=\exp(-S)$.


\section{Results and discussion}
\label{sec:results}

We systematically studied the complexes of vacancies and oxygen. We built up the following defect models: (i) carbon-vacancy (V) which has four neighbor atoms. A neighbor atom can be either a carbon atom with a dangling bond or oxygen substituting that carbon atom; (ii) oxygen substituting carbon atom (O) which may form different number of bonds depending on its position (e.g., near the vacancy). We model the defects by varying their distance along $\langle 100 \rangle$ direction within (110) plane with assuming $C_{1h}$ symmetry configurations. The created defects are labeled as OV, OCV, OCCV, OCCCV, and OCCCCV. In OV defect, the vacancy is the first nearest neighbor of the substitutional oxygen defect. In the OCV, OCCV, and OCCCV defect, the vacancy lies in the second, third, and fourth nearest neighbor of substitutional oxygen defect, respectively. We did not consider larger separation of the oxygen and vacancy because fast screening by PBE calculations in a giant supercell indicates that the formation energy increases by enlarging the distance between the two species.

We note that the OV defect has been already characterized and associated with WAR5 ESR center~\cite{Thiering2016} but we show the results for this defect for the sake of completeness and for comparison to show the general trends.

After optimization of defects using HSE06 functional, the geometry of defects varies with the distance between the oxygen atom and the vacancy (see Fig.~\ref{fig:structures}). In the OCV defect, the carbon-oxygen bond still exists since the carbon has a dangling bond near the vacant site that can form a bond with the oxygen atom. However, in OCCV and OCCCV defects, the oxygen cannot form a bond with the carbon dangling bond from the vacancy. Rather new types of bonds are developed between the carbon atoms at the vacant site and the nearest neighbor carbon atom of the oxygen defect as the oxygen atom prefers to form three bonds in diamond with donating one electron to the dangling bond of the vacant site. We note that an exotic carbon cluster forms in the OCCCV defect with consisting of sp$^2$ carbon atoms in the intersection of the vacant site and the oxygen atom. \\

In the forthcoming sections, we describe the electronic structure of these vacancy-oxygen complexes and determine the formation energies in their various charge states with identifying the most stable defect structures. We continue with the computation of the key magneto-optical parameters and spectrum for the relevant defect structures with short discussion in the context of defect qubits in diamond. 

\begin{figure*}[ht!]
  \includegraphics[scale=.8]{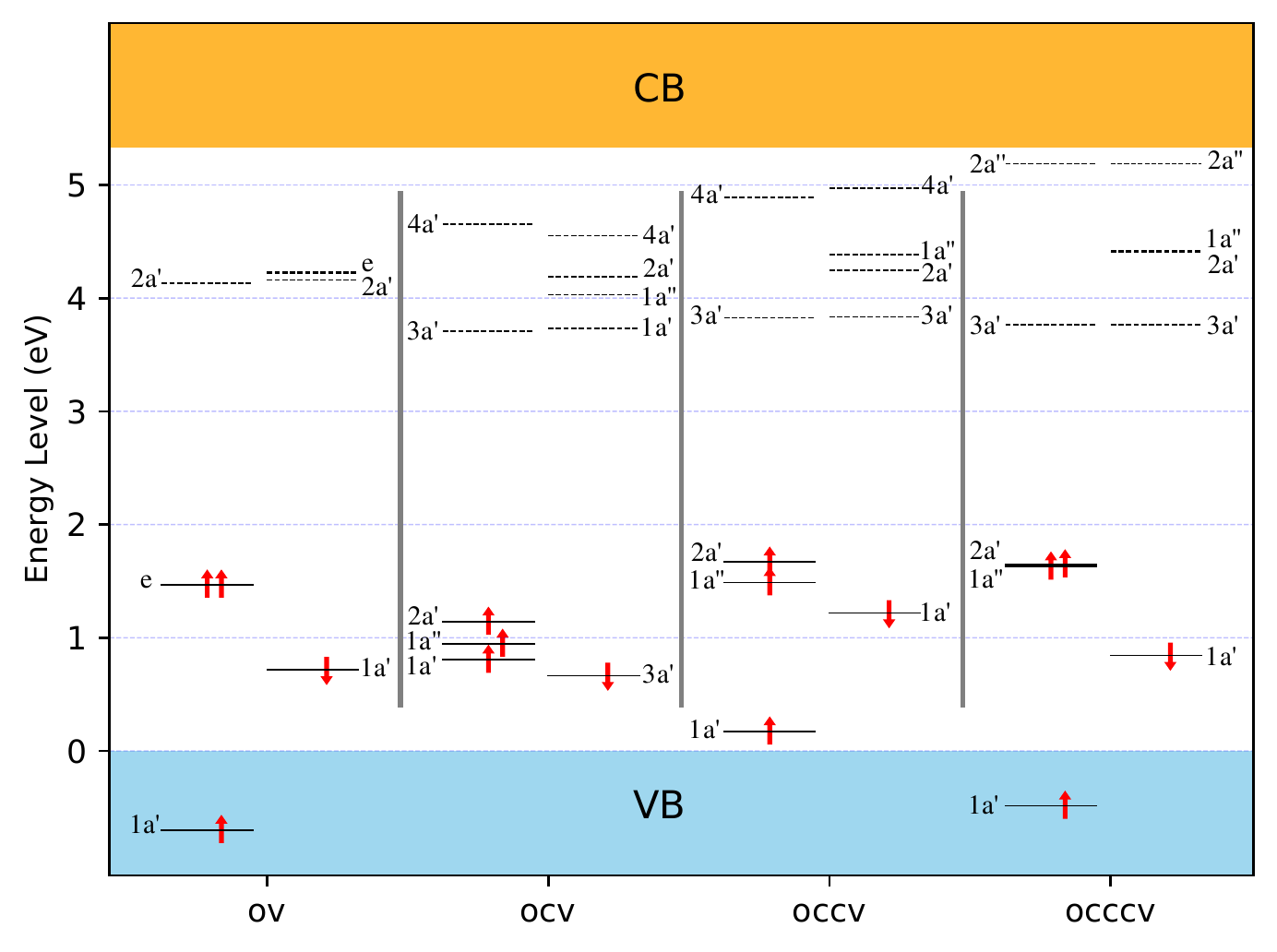}
    \caption{\label{fig:kshse}
   Electronic structure for the ground states of the neutral defects using HSE06 functional. Kohn-Sham (KS) levels are represented by spin-up ($\uparrow$) and spin-down ($\downarrow$). The valence band (VB) and conduction band (CB) are shown in cyan and orange colors, respectively.}
\end{figure*}
\subsection{Electronic structure of defects}

The electronic structure of defects in the ground state is depicted in Fig.~\ref{fig:kshse}.
In the neutral OV, oxygen forms three bonds to carbon atoms and leave three carbon dangling bonds behind in C$_{3v}$ symmetry. The three carbon dangling bonds will introduce a low-lying $a_1$ level and a higher-lying double degenerate $e$ level occupied by four electrons. Furthermore, an empty $a_1$ level also appears close to the conduction band minimum (CBM), which is an antibonding orbital localized to the oxygen-carbon bonds. As previously discussed~\cite{Thiering2016} the OV defect with $C_{3v}$ symmetry with $S=1$ triplet ground state is very similar to that of NV$^-$ as two electrons will occupy the double degenerate $e$ level.

Next, we discuss the electronic structure of the other considered oxygen-vacancy complexes.

In OCV six dangling bonds occur. Four dangling bonds originate from the vacant site and two ones from the oxygen site. The spin-up occupied states are mainly localized on the four carbon dangling bonds around the vacant site but the spin-down occupied state is localized on the oxygen and carbon dangling bonds at the oxygen site. The empty states are localized on the carbon-oxygen antibonding orbital and the carbon dangling bond at the oxygen site [see Fig.~ \ref{fig:structures}(a)]. We obtained KS wavefunctions with broken symmetry, i.e., the KS wavefunctions do not show the $C_{1h}$ symmetry, which is a signature of a highly correlated electronic system as explained for other dangling bond defects in diamond in Ref.~\onlinecite{Thiering2016}. In the majority spin-up channel, all of the KS states originate from the same type of carbon dangling bond with resulting in quasidegenerate levels slightly split by the low-symmetry crystal field of the defect. All the quasidegenerate levels are occupied by electrons in the neutral charge state with resulting in a $S=1$ high-spin ground state. The occupied and empty levels indicate that the defect may exist in multiple charge states depending on the position of the Fermi-level.

OCCV defect has six dangling bonds. Four dangling bonds originate from the vacant site and two ones from the oxygen site. The occupied states are mainly localized on the five carbon dangling bonds (one from oxygen site), and the empty states are localized on the carbon-oxygen antibonding orbital and the carbon dangling bond at the oxygen site [see Fig.~\ref{fig:structures}(b)]. Again, quasidegenerate orbitals appear that are introduced by the carbon dangling bonds and occupied by two electrons with stabilizing the $S=1$ high-spin ground state. The KS wavefunctions again exhibit broken symmetry solution so it has a highly correlated electronic solution. The defect may exist in other charge states too depending on the position of the Fermi-level.
 
OCCCV defect exhibits a specific structure. Because of the formation of the carbon cluster in the middle part of the defect sp$^2$ orbitals appear that produce defect levels in the fundamental band gap beside the levels originating from the three sp$^3$ carbon dangling bonds around the vacant site. The occupied states are mainly localized on the five carbon atoms in the defect, and the empty states are localized on the carbon-oxygen antibonding orbital and the carbon dangling bond at the oxygen site [see Fig.~\ref{fig:structures}(c)]. Again, quasidegenerate levels appear in the gap that is occupied by two electrons that results in $S=1$ high-spin ground state.

We conclude that the considered oxygen-vacancy complexes realize $S=1$ high-spin systems that might act as quantum bits. On the other hand, none of these defects could be associated with the ST1 ODMR center, which has $S=0$ singlet ground state. Nevertheless, it is worthy to further study these complexes that are paramagnetic center that may realize qubits. These depends on their structural and photoexcitation stabilities. We investigate these issues by computing the formation energy of these defects in various charge states.
\begin{figure*}[ht]
    \centering
    \includegraphics[scale=0.14]{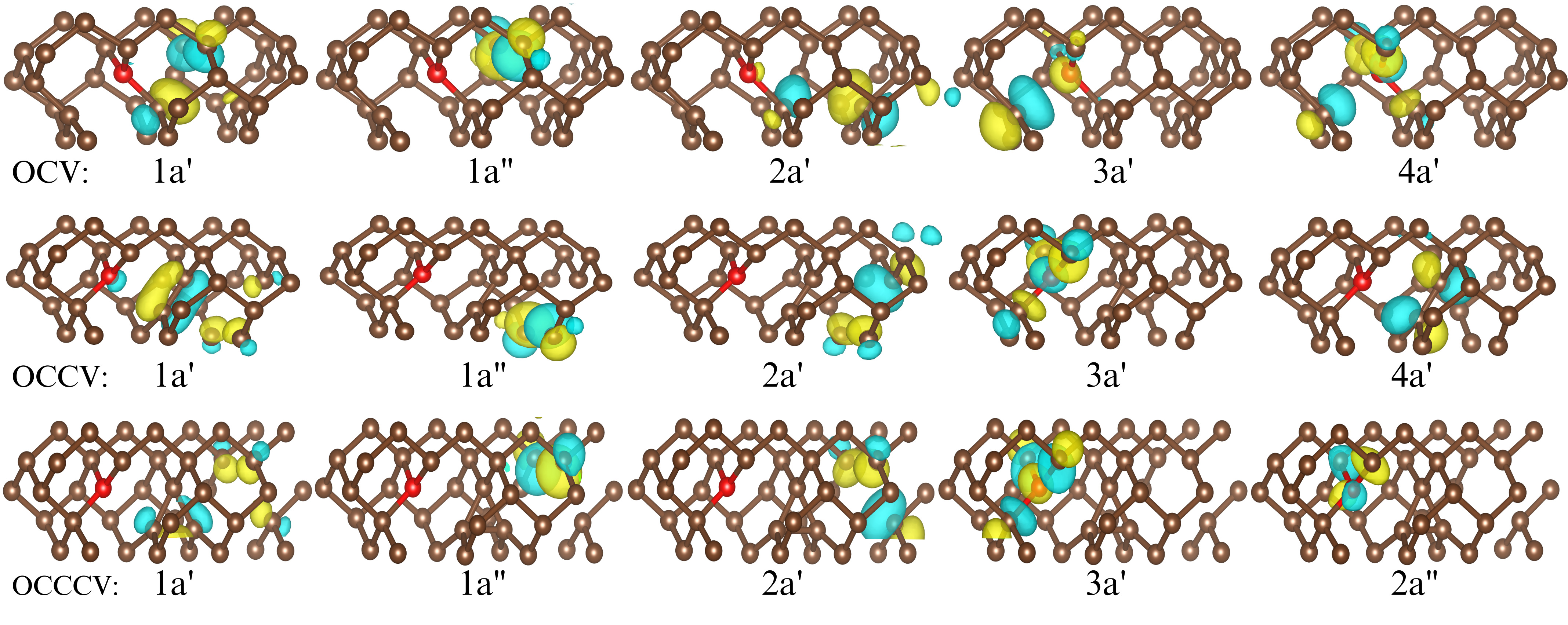}
    \caption{\label{fig:wfc}
   Representation of Kohn-Sham orbitals of the OCV, OCCV and OCCCV defects. The light cyan and yellow lobes exhibit negative and positive isovalues. The isosurface absolute value is set to $5.5 \times 10^{-7}$ {\AA}$^{-3}$. }
\end{figure*}

\subsection{Formation energy of defects}

\begin{figure}[b]
\includegraphics[width=1 \linewidth]{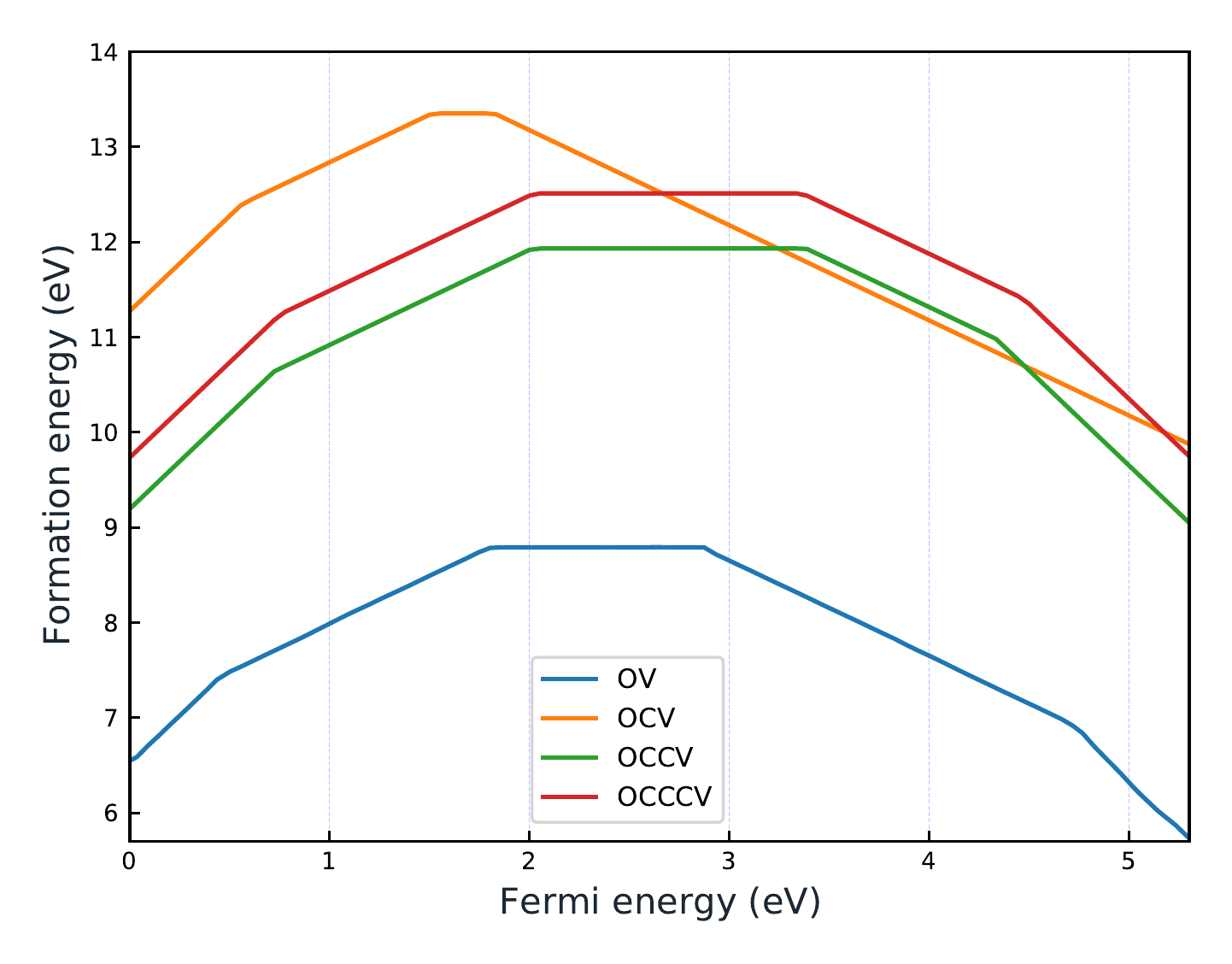}
\caption{\label{fig:formation}Formation energy of the defects as a function of the position of the Fermi-level. The valence band maximum (VBM) is aligned to zero for the sake of simplicity. The conduction band minimum (CBM) is at 5.4~eV. For each defect the ($2+$), ($+$), ($0$), ($-$) and ($2-$) charge states may appear except for the OCV defect, which is not stable in the ($2-$) charge state. The ($+|0$) and ($0|-$) charge transition levels of OCCV defect are at VBM$+2.0$~eV and VBM$+3.3$~eV (CBM$-2.1$~eV), respectively, which are at the respective crossing points of the green curve.}
\end{figure}

The defects formation energy in this study after optimization are depicted in Fig.~\ref{fig:formation}. 
By comparing the formation energies of the same type of defects with different configurations, we recognize a non-trivial trend for defects. The OV defect is the most stable defect. The OCV defect has much higher formation energy than the OV defect. However, the formation energy of OCCV defect lies between the formation energies of OV and OCV defects and then the formation energy of OCCCV defect is again higher than that of OCCV defect. This means that OCCV defect is a metastable configuration.

We note that formation of complex may occur with combination of mobile vacancy and the immobile substitutional oxygen. The probability of the reaction can be quantified by the binding energy. In our definition, the positive value of the binding energy indicates the favor of formation of the complex from the constituting species. In the neutral charge state, the calculated binding energies are 5.33 and 2.57 eV for OV and OCCV defects, respectively, and they are 1.29 and 2.26 eV for OCV and OCCCV defects, respectively. This indicates that OV and OCCV defects are the most stable species and they have a higher probability to form rather than other defects. The metastable OCCV defect may exist with using an appropriate annealing procedure for the diffusing vacancies. This phenomena has been recognized already for the adjacent vacancies configurations, i.e., divacancy in diamond~\cite{slepetz2014}, where VCCV was found to be a metastable complex. We think that a very similar scenario shows up for the OCCV defect.

It is apparent from Fig.~\ref{fig:kshse} that all the considered defects may be ionized by either removing electrons from the filled levels or adding electrons to the empty levels. We calculated the formation energies of the defects in their various charge states as a function of the position of the Fermi-level (see Fig.~\ref{fig:formation}). The charge transition levels can be recognized at the crossing points between formation energy curves of the given charge states for each defect. It can be observed that all the considered defects but OCV defect are stable in their neutral charge state for the Fermi-level position at the middle of the gap, which is the most likely quasi Fermi-level of implanted diamond. The OCV defect exists in its negative charge state but it is still the least stable complex among the considered ones. Again, the main conclusion is that the OCCV defect is a metastable complex.

The photoionization threshold energies can be read from the plot as the charge transition level with respect to either the conduction band minimum (positive ionization) or the valence band maximum (negative ionization). It can be concluded from the calculated charge transition levels that the neutral charge state of the OV and OCCV defects is photostable against 532-nm (2.33-eV) photoexcitation: the neutral state cannot be ionized either to conduction band minimum nor from the valence band maximum with green illumination. Rather green illumination would continuously cycle the OCCV defect to the neutral charge state under any doping condition as an electron from the valence band would be promoted to the empty defect level in the single positive charge state or the electron would be promoted from the highest occupied defect level to the conduction band in the single negative charge state (see Fig.~\ref{fig:formation}). Only ultraviolet exposure can temporarily drive out the OCCV defect from the neutral charge state. 

We finally conclude that beside the already known OV defect the OCCV defect may exist in diamond as the complex oxygen and vacancy, and the OCCV defect can be observed in its neutral charge under typical measurement conditions. Therefore we apply our magneto-optical characterization techniques mostly but not exclusively on OCCV defect in the next sections.

\subsection{Zero-field-splitting} 

One of the most characteristic magneto-optical parameters of paramagnetic centers is the zero-field-splitting (ZFS). As the considered defects consist of only light atoms we assume that the ZFS originates from the electron spin-spin dipolar interaction. Although, the most important defect in this study is the OCCV defect, nevertheless, we provide the previously computed ZFS $D$ constant for OV defect~\cite{Thiering2016} and our calculated ZFS $D$ and $E$ constants for OCV and OCCCV defects too. The results are listed in Table~\ref{tab:tablehse} together with the relevant experimental data for comparison. The motivation behind showing the results for all of these defects is to explore whether is there any trend on the ZFS constants with enlarging the separation distance between the substitutional oxygen and the vacancy in oxygen-vacancy complexes. We note that it was an assumption in an abstract modeling of the ST1 defect~\cite{Lee2013} that separation of the vacancy from the core of the defect at larger distances may reduce the $D$ constant. 

We start the discussion with the OV defect. It is a special case because the closest pair of oxygen and vacancy has the highest $C_{3v}$ symmetry and the electronic structure of the ground state is similar to that of NV$^-$ center. The spin densities are localized on the three carbon dangling bonds of the vacancy within the second neighbor distance in diamond. This results in about 2.9~GHz $D$ constant and the orthorhombic $E$ constant is zero because of the high $C_{3v}$ symmetry of the defect. As the vacancy placed farther from the position of the oxygen the symmetry is reduced to $C_{1h}$, thus the orthorhombic $E$ parameter emerges. In the OCCCV defect the $E$ parameter is reduced compared to that of OCV and OCCV defects. We explain this feature by the fact that the spin densities are exclusively localized on the sp$^3$ dangling bonds of the remote vacancy (see $1a^{\prime\prime}$ and $2a^{\prime}$ orbitals in Fig.~\ref{fig:wfc}) that are almost equally far from each other. It is expected that the $E$ constant would further reduce at larger separation of O and V. On the other hand, the $D$ constant fluctuates around 2.8~GHz in all the considered oxygen-vacancy complexes and does not show any significant reduction when the separation between O and V is enlarged. The reason behind this observation is that the spin density always distributed in the vacant site of the complex with sp$^3$ dangling bonds within the second neighbor distance as indicated by the spinpolarized $1a^{\prime\prime}$ and $2a^{\prime}$ orbitals in OCV, OCCV and OCCCV defects (Fig.~\ref{fig:wfc}). 

It is obvious that neither the spin-state of the electronic ground state nor the calculated ZFS parameters of the considered oxygen-vacancy complexes can be associated with the ST1 ODMR center, and the fine details in the electronic structure could be decisive in the value of the $D$ constant that may significantly differ from that of simple models based on the distance between sp$^3$ dangling bonds.
\begin{table}[hbt!]
\caption{\label{tab:tablehse}
The formation energy ($E_\text{f}$), the spin state of the electronic ground state, the symmetry of the defect structure, and the ZFS $D$ and $E$ parameters of the triplet state for each neutral defect are listed. The relevant experimental data for WAR5 ESR center~\cite{Breeze2016} and ST1 ODMR center~\cite{Lee2013} are also shown for comparison.}
\begin{ruledtabular}
\begin{tabular}{cccccc}
 Defect & $E_\text{f}$(eV) & GS & Symmetry &
 $D$ (MHz) & $E$ (MHz)  \\
\hline 
ST1(exp.) &-& S & C$_{2v}$ or lower& 1139 & 135  \\ 
WAR5 (exp.) &-& T & C$_{3v}$ & 2888 & 0 \\
OV   & 7.79  & T & C$_{3v}$ & 2989 & 0    \\ 
OCV & 13.35 & T & C$_{1h}$ & 2886 & 160 \\ 
OCCV & 11.93 & T & C$_{1h}$ & 2771 & 433 \\ 
OCCCV& 12.51 & T & C$_{1h}$ & 2929 & 44 \\ 
\end{tabular}
\end{ruledtabular}
\end{table}

\subsection{Hyperfine interaction}

\begin{figure}[h!]
  \includegraphics[width=1 \linewidth]{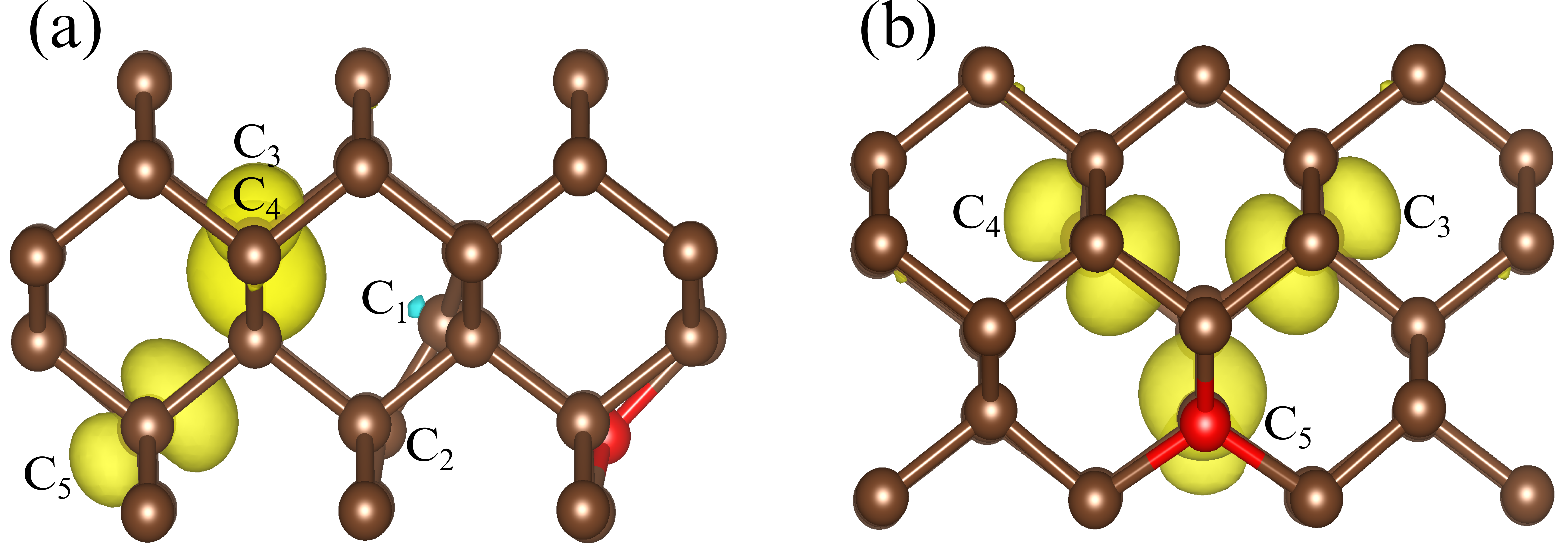}
\caption{\label{fig:spin} Spin density distribution around the core of the OCCV defect with two different views: (a) (1$\bar{1}$0) plane and (b) close to (110) plane. O and C atoms are shown by red and brown spheres, respectively. The hyperfine constants for the atoms labeled here are given in Table~\ref{tab:tablehyper}. Yellow lobes exhibit positive isovalues of the calculated spin density.}
\end{figure}

We now focus on the properties of the metastable OCCV defect. The defect has paramagnetic ground state that may be observed by ESR techniques. We already provided the ZFS constants above. We continue with the computation of hyperfine tensors that reflects the interaction between the electron spin and nuclear spins in the lattice. The observed hyperfine lines in the ESR spectrum can yield important information about which types of atoms are
involved in the defect, the overall symmetry of the defect is, and upon which atoms the electron spin density $n(r)$ is primarily located (e.g., Ref.~\onlinecite{Kriszti2013}).‌ In other words, the ESR method directly provides information about the chemical composition of the defects if nuclear spins are present in the core of the defect. This principle has been recently applied to identify an oxygen-vacancy complex in hexagonal boron nitride~\cite{Song2022}. Thus, we computed the hyperfine tensors for neutral OCCV defect for future identification. If the defect is created by oxygen implantation then $^{17}$O isotopes may be used with $I=1/2$ nuclear spins. Furthermore, $^{13}$C nuclear spins may appear near the defect with 1.1\% natural abundance in the diamond lattice. The calculated spin density is shown in Fig.~\ref{fig:spin} where we label the atoms for which the hyperfine tensors were computed. Because of the $C_{1h}$ symmetry, the spin density so the hyperfine constants on the C$_3$ and C$_4$ atoms are the same. Small spin density can be found on the C$_1$, C$_2$, and the oxygen atoms whereas the majority of the spin density is localized on the C$_{3-5}$ sp$^3$ dangling bonds. After diagonalizing the computed hyperfine tensors the $A_{xx}$, $A_{yy}$ and $A_{zz}$ hyperfine constants are obtained where $A_{zz}$ is chosen to be the largest absolute value by convention. We list the the computed hyperfine constants in Table~\ref{tab:tablehyper}.

\begin{table}[hbt!]
\caption{\label{tab:tablehyper}
Hyperfine constants (A$_{xx}$,
A$_{yy}$ and A$_{zz}$ ) for the nearest-neighbor atoms for the OCCV defects. The atom labels are shown in Fig.~\ref{fig:spin}. The reported values are given together with the computed core spinpolarization.}
\begin{ruledtabular}
\begin{tabular}{ccccc}
 Atoms & A$_{xx}$ (MHz) & A$_{yy}$ (MHz) & 
 A$_{zz}$ (MHz)  \\
\hline 
O       & 7.7 & 7.6 & 8.1  \\
C$_{1}$ & -2 & -1.4 & -10  \\
C$_{2}$ & 3.7 & 3.6 & 12.5 \\
C$_{3,4}$ & 107.5 & 107.2 & 197 \\
C$_{5}$ & 109.4 & 109.3 & 180.5 \\
\end{tabular}
\end{ruledtabular}
\end{table}

\subsection{Photoluminescence spectrum}

It can be revealed from Fig.~\ref{fig:kshse} that the electron from the in-gap $2a^\prime$ state to the $3a^\prime$ state in the OCCV defect may decay radiatively back to the ground state. Phonons may assist the optical transition. The contribution of the phonon sideband may be estimated by the magnitude of the change of the coordinates ($\Delta Q$) upon optical excitation within Huang-Rhys theory, which is defined in Eq.~\ref{eq:deltaq}. The OCCV defect exhibits $\Delta Q = 0.94$ for the optical transition described above, which implies a large phonon sideband in the optical spectrum. 

\begin{figure}[h!]
  \includegraphics[width=0.8\linewidth]{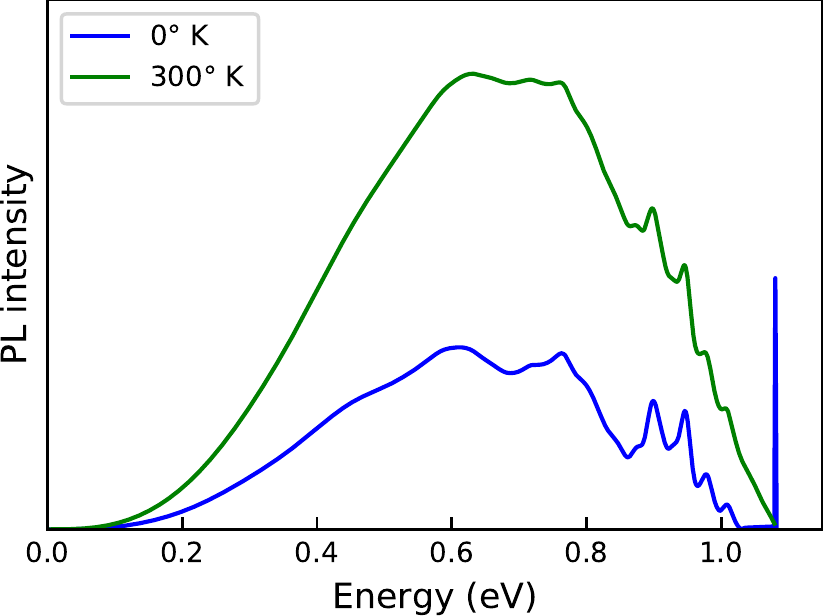}
\caption{\label{fig:PL} The calculated fluorescence spectrum of the OCCV defect at $T=0$~K and at room temperature (green line). A gaussian broadening of $2$~meV was applied for the ZPL line.}
\end{figure}

The final fluorescence spectrum of the neutral OCCV defect as computed within Huang-Rhys theory is plotted in Fig.~\ref{fig:PL}. The calculated ZPL energy is at 1.08~eV. The calculated $S=7.33$ corresponds to $W=0.06\%$, thus the ZPL emission is weak. In the simulated room temperature spectrum it becomes almost invisible. The phonon sideband is indeed very wide and it stretches down to 0.1~eV. 
The OCCV defect is a near-infrared emitter, which is rare in diamond. Recently, a telecom emitter has been reported in diamond~\cite{Mukherjee2022}. In that emitter, the nonradiative decay is very efficient. It is likely that the same effect occurs for the OCCV defect and the emission would be weak.

\section{Conclusions}
\label{sec:conclusion}

We systematically studied the electronic structure, formation energies, and magneto-optical parameters of complexes of oxygen and vacancy in diamond by means of density functional theory plane wave supercell calculations. We found that these defects introduce multiple levels into the fundamental band gap. The defect states are highly localized on the oxygen, oxygen-carbon bonds and the carbon dangling bonds in the core of the defect complexes. We showed that the ground state should be calculated with advanced DFT functionals for accurate prediction because the highly localized dangling bonds are strongly interacting with each other and produce high correlation among the orbitals. We found that these oxygen-vacancy complexes have a high-spin $S=1$ ground state. The OCCV complex is a metastable defect that may occur in oxygen doped diamond. The OCCV complex is a near-infrared color center with specific ZFS and hyperfine constants that might be observed in photoluminescence and electron spin resonance techniques in oxygen doped diamond.

\begin{acknowledgments}
Support by the National Excellence Program for the project of Quantum-coherent materials (NKFIH Grant No.\ KKP129866) as well as by the Ministry of Culture and Innovation and the National Research, Development and Innovation Office within the Quantum Information National Laboratory of Hungary (Grant No.\ 2022-2.1.1-NL-2022-00004), and by the European Commission for the project QuMicro (Grant No.~) as well as the QuantERA II project MAESTRO is much appreciated. We acknowledge the high-performance computational resources provided by KIF\''U (Governmental Agency for IT Development) institute of Hungary. Open access funding provided by ELKH Wigner Research Centre for Physics.
\end{acknowledgments}

\bibliography{bibliography}
\end{document}